\documentclass[english,10pt,aps,prd,a4paper,preprintnumbers,floatfix,nofootinbib,showpacs,superscriptaddress]{revtex4-1}
\usepackage[english]{babel}
\usepackage{amsmath}
\usepackage{graphicx}
\usepackage{color}
\usepackage{mathrsfs}   
\usepackage{amssymb}
\usepackage{hyperref}
\usepackage[normalem]{ulem} 
\usepackage{dcolumn}
\usepackage{bm}
\usepackage{graphicx}
\usepackage[sort&compress]{natbib}

\newcommand {\be} {\begin{equation}}
\newcommand {\ee} {\end{equation}}

\newcommand*\rfrac[2]{{}^{#1}\!/_{#2}}

\definecolor{greenLinks}{rgb}{0, 0.6, 0} 
\definecolor{blueLinks}{rgb}{0, 0, 0.6}
\definecolor{redLinks}{rgb}{0.6, 0, 0}
\definecolor{tempText}{rgb}{0.55, 0.10,0.67}
\definecolor{eprintLinks}{rgb}{0.4, 0.4, 0.4}
\definecolor{journalLinks}{rgb}{0.6, 0, 0}

\newcommand{\MYhref}[3][redLinks]{\href{#2}{\color{#1}{#3}}}%

\def\vev#1{\left\langle #1\right\rangle}

\usepackage{float}

\def\vev#1{\left\langle #1\right\rangle}

\def\21{$\mathrm{SU(2)_L \otimes U(1)_Y}$ }
\def\31{$\mathrm{SU(3)_c \otimes U(1)_Q}$ }
\def\SM{$\mathrm{SU(3)_c \otimes SU(2)_L \otimes U(1)_Y}$ }

\def\3211{$\mathrm{SU(3) \otimes SU(2)_L \otimes U(1)_R \otimes U(1)_{B-L}}$ }
\def\321{$\mathrm{SU(3) \otimes SU(2) \otimes U(1)}$ }
\def\422{$\mathrm{SU(4) \otimes SU(2) \otimes SU(2)_R}$ }
\newcommand {\ignore}[1]{}

\newcommand{\sm}{{Standard Model }}

\def\vev#1{\left\langle #1\right\rangle}

\def\SM{$\mathrm{ SU(3)_C \otimes SU(2)_L \otimes U(1)_Y }$ }

\newcommand{\AddrAHEP}{%
  AHEP Group, Institut de F\'{i}sica Corpuscular --
  C.S.I.C./Universitat de Val\`{e}ncia, Parc Cient\'ific de Paterna.\\
 C/ Catedr\'atico Jos\'e Beltr\'an, 2 E-46980 Paterna (Valencia) - SPAIN}

\begin{document}

 \title{Generalized Bottom-Tau unification, neutrino oscillations and dark matter:\\
 predictions from a lepton quarticity flavor approach  }

\author{Salvador Centelles Chuli\'{a}}\email{salcen@ific.uv.es}
\affiliation{\AddrAHEP}
\author{Rahul Srivastava}\email{rahulsri@ific.uv.es}
\affiliation{\AddrAHEP}
\author{Jos\'{e} W. F. Valle}\email{valle@ific.uv.es}
\affiliation{\AddrAHEP}

\begin{abstract}
   \vspace{1cm}
   
   We propose an $A_4$ extension of the \sm with 
   a Lepton Quarticity symmetry correlating dark matter stability with
   the Dirac nature of neutrinos. The flavor symmetry predicts
   (i) a generalized bottom-tau mass relation involving all families,
   (ii) small neutrino masses are induced \textit{a la seesaw}, 
   (iii) CP must be significantly violated in neutrino oscillations,  
   (iv) the atmospheric angle $\theta_{23}$ lies in the second octant, and
   (v) only the normal neutrino mass ordering is realized.

\end{abstract}

\maketitle


\section{Introduction}
\label{sec:introduction}

Probably the number one mystery in particle physics is the
understanding of the pattern of fermion masses and mixings from first
principles. Indeed, the charged fermion mass pattern is not described
in the theory: the \sm only allows us the freedom to fit the observed
charged fermion masses, while lacking the masses of neutrinos
altogether.
An approach towards addressing, at least partially, the charged
fermion mass problem, is the possibility of relating quarks and lepton
masses as a result of a flavor symmetry~\cite{Morisi:2011pt}, i.e.
\begin{equation}
  \label{eq:b-tau}
\frac{m_b}{{\sqrt{m_d m_s}}} = \frac{m_\tau}{{\sqrt{m_e m_\mu}}}~.
\end{equation}
Notice that this mass relation constitutes a consistent
flavor-dependent generalization of the conventional bottom-tau SU(5)
prediction, but does not require grand-unification. It provides a
partial solution to the charged fermion mass problem, which can be
shown to hold in some theories of flavor based on the
$A_4$~\cite{Morisi:2011pt,King:2013hj,Morisi:2013eca} and
$T_7$~\cite{Bonilla:2014xla} symmetries.

Turning to neutrinos, the origin of their mass, the understanding of
their mixing properties and the puzzle of whether they are their own
anti-particles continue to defy theorists.  Underpinning the solution
to such neutrino puzzles may not only write a new chapter of particle
physics, but also shed light on astrophysical and cosmological
puzzles.
One of the latter is the puzzle of Dark Matter, believed to be
associated to the existence of a new absolutely or nearly stable
neutral particle.

There have been attempts at formulating joint solutions to the above
shortcomings of the standard model.
For example, in scotogenic models dark matter is introduced as a
messenger of radiative neutrino mass
generation~\cite{Ma:2006km,Hirsch:2013ola,Merle:2016scw} whose
stability follows from the radiative nature of the neutrino mass.
Several alternative ideas have come out, invoking non-Abelian flavor
symmetries~\cite{Hirsch:2012ym,Morisi:2012fg,King:2014nza}, such as
the $A_4$ symmetry~\cite{Ma:2001dn,Babu:2002dz,Morisi:2013qna}.
For example, dark matter could be stable as a result of some remnant
of the flavor symmetry associated to the pattern of neutrino
mixing~\cite{Hirsch:2010ru,Boucenna:2012qb}.
In all these models neutrinos are Majorana type. 

However, recently there has been a renewed interest in Dirac
neutrinos~\cite{Chulia:2016ngi, Chulia:2016giq, Aranda:2013gga,
  Bonilla:2016diq, Ma:2016mwh, Ma:2015raa, Ma:2015mjd,
  Bonilla:2016zef, Valle:2016kyz, Abbas:2016qbl, Abbas:2013uqh,
  Wang:2016lve, Wang:2017mcy, Okada:2014vla, Borah:2016zbd,
  Borah:2017leo} which may attain naturally small masses in many
scenarios. 
For example, it can happen that the same flavor symmetry which
presumably sheds light on the pattern of neutrino oscillation
parameters, also implies that neutrinos are Dirac
fermions~\cite{Aranda:2013gga}.
Specially tantalizing is the idea that the stability of dark matter
can be directly traced to the Dirac nature of neutrinos~\cite{
  Chulia:2016ngi, Chulia:2016giq, Bonilla:2016diq, Ma:2016mwh}.
One way to realize this idea is by means of a $Z_4$ Lepton Quarticity
symmetry~\cite{ Chulia:2016ngi, Chulia:2016giq}. Within such approach
the same $Z_4$ discrete lepton number symmetry ensures the stability
of dark matter and the absence of all the Majorana mass terms. Thus
owing to Lepton Quarticity, the Dirac nature of neutrinos and the
stability of dark matter are intimately related: the breakdown of this
symmetry will simultaneously imply loss of dark matter stability as
well as the \textit{Diracness} of neutrinos.
 
Here we focus on the Lepton Quarticity models of dark matter, along
the lines pursued in~\cite{ Chulia:2016ngi,Chulia:2016giq}.
The plan of the paper is as follows.  In Sect.~\ref{sec:model} we
sketch in some detail the extended particle content required to
realize the non-Abelian flavor symmetry of the model, and show how the
Dirac nature of neutrinos and the smallness of their seesaw--induced
masses both follow from our non-Abelian discrete flavor symmetry. We
also briefly discuss the appearance of a viable WIMP dark matter
candidate in this model.
In Sect.~\ref{sec:numerical-scan} we present our predictions for the
current and future neutrino oscillation experiments.
We find that the atmospheric angle $\theta_{23}$ and the CP phase
$\delta_{CP}$, whose current experimental determination is still
rather poor, are tightly related to each other within our model.
Finally we summarize our results in
Sect.~\ref{sec:summary-conclusions-}.


 \section{The model setup }
\label{sec:model}


Here we describe the model in some detail.  The particle content of
our model along with the $SU(2)_L \otimes Z_4 \otimes A_4$ charge
assignments of the particles are given in Table \ref{tab1}.
\begin{table}[ht]
\begin{center}
\begin{tabular}{c c c c || c c c c}
  \hline \hline
Fields          \hspace{1cm}    & $SU(2)_L$           \hspace{1cm}   & $A_4$           \hspace{1cm}   &  $Z_4$	    \hspace{1cm}    & Fields          \hspace{1cm}    & $SU(2)_L$           \hspace{1cm}   & $A_4$           \hspace{1cm}   &  $Z_4$	     \\
\hline \hline
$\bar{L}_i$     \hspace{1cm}    & $\mathbf{2}$    \hspace{1cm}    & $\mathbf{3}$    \hspace{1cm}   &  $\mathbf{z}^3$   \hspace{1cm}   &		
$\nu_{e,R}$     \hspace{1cm}    & $\mathbf{1}$    \hspace{1cm}    & $\mathbf{1}$    \hspace{1cm}   &  $\mathbf{z}$      \\
$\bar{N}_{i,L}$ \hspace{1cm}    & $\mathbf{1}$    \hspace{1cm}    & $\mathbf{3}$    \hspace{1cm}   &  $\mathbf{z}^3$   \hspace{1cm}   &        
$\nu_{\mu, R}$  \hspace{1cm}    & $\mathbf{1}$    \hspace{1cm}    & $\mathbf{1'}$   \hspace{1cm}   &  $\mathbf{z}$       \\
$N_{i,R}$       \hspace{1cm}    & $\mathbf{1}$    \hspace{1cm}    & $\mathbf{3}$    \hspace{1cm}   &  $\mathbf{z}$      \hspace{1cm}  &		
$ \nu_{\tau,R}$ \hspace{1cm}    & $\mathbf{1}$    \hspace{1cm}    & $\mathbf{1''}$  \hspace{1cm}   &  $\mathbf{z}$        \\
$ l_{i,R} $     \hspace{1cm}    & $\mathbf{1}$    \hspace{1cm}    & $\mathbf{3}$    \hspace{1cm}   &  $\mathbf{z}$      \hspace{1cm}  &		
$ d_{i,R} $     \hspace{1cm}    & $\mathbf{1}$    \hspace{1cm}    & $\mathbf{3}$    \hspace{1cm}   &  $\mathbf{z}$        \\
$ \bar{Q}_{i,L} $  \hspace{1cm}    & $\mathbf{2}$    \hspace{1cm}    & $\mathbf{3}$    \hspace{1cm}   &  $\mathbf{z}^3$    \hspace{1cm}  &		
$ u_{i,R} $     \hspace{1cm}    & $\mathbf{1}$    \hspace{1cm}    & $\mathbf{3}$    \hspace{1cm}   &  $\mathbf{z}$       	\\
\hline
$\Phi_1^u$    \hspace{1cm}        & $\mathbf{2}$      \hspace{1cm}        & $\mathbf{1}$      \hspace{1cm}  	    &  $\mathbf{1}$        \hspace{1cm} &		
$\chi_i$        \hspace{1cm}        & $\mathbf{1}$      \hspace{1cm}   	    & $\mathbf{3}$    	\hspace{1cm} 	    &  $\mathbf{1}$            \\
$\Phi_2^u$    \hspace{1cm}        & $\mathbf{2}$      \hspace{1cm}        & $\mathbf{1'}$     \hspace{1cm}        &  $\mathbf{1}$        \hspace{1cm}     &		             
$\eta$          \hspace{1cm}        & $\mathbf{1}$	\hspace{1cm}        & $\mathbf{1}$      \hspace{1cm}        &  $\mathbf{z}^2$      	      \\
$\Phi_3^u$    \hspace{1cm}        & $\mathbf{2}$	\hspace{1cm}        & $\mathbf{1''}$    \hspace{1cm}        &  $\mathbf{1}$        \hspace{1cm}     &		
$\zeta$         \hspace{1cm}        & $\mathbf{1}$	\hspace{1cm}        & $\mathbf{1}$      \hspace{1cm}        &  $\mathbf{z}$       	 \\
$\Phi_i^d$      \hspace{1cm}        & $\mathbf{2}$	\hspace{1cm}        & $\mathbf{3}$      \hspace{1cm}        &  $\mathbf{1}$        \hspace{1cm}     &		           
                \hspace{1cm}        & 		 	\hspace{1cm}        & 			\hspace{1cm}        &     	          \\
    \hline
  \end{tabular}
\end{center}
\caption{Charge assignments for leptons, quarks,  scalars 
  ($\Phi_i^u$, $\Phi_i^d$ and $\chi_i$) as well as ``dark matter sector'' 
  ($\zeta$ and $\eta$). Here $\mathbf{z}$ is the fourth root
  of unity, i.e. $\mathbf{z}^4 = 1$.  }
 \label{tab1} 
\end{table}
Note that in Table~\ref{tab1} the $L_i = (\nu_i, l_i)^T$,
$i = e, \mu, \tau$ denote the lepton doublets, transforming as
indicated under the flavor symmetry.

Apart from the \sm fermions, the model also includes three
right--handed neutrinos $\nu_{i,R}$ which are singlets under the \SM
gauge group, singlets under $A_4$, but carry charge $\mathbf{z}$ under
$Z_4$. We also add three gauge singlet Dirac fermions
$N_{i,L}, N_{i,R}$; $i = 1, 2, 3$ transforming as triplets of $A_4$
and with charge $\mathbf{z}$ under $Z_4$, as shown in
Table~\ref{tab1}.
Notice that in the scalar sector we have two different sets of fields
$\Phi^u_i, \Phi^d_i$; $i = 1,2,3$, which are all doublets under the
SU(2)$_L$ gauge group, both sets transforming trivially under
$Z_4$. Under the $A_4$ flavor symmetry, $\Phi_i^d$ transforms as a
triplet, while $\Phi_i^u$ transform as singlets. In addition to the
above symmetries we also impose an additional $Z_2$
symmetry\footnote{This additional $Z_2$ symmetry is only required in a
  non-supersymmetric variant. Clearly the model can be easily
  supersymmetrized, in which case this additional $Z_2$ symmetry is no
  longer required.}.  Under this $Z_2$ symmetry, all the fields
transform as $1$ except for $\Phi^d_i$, $l_{i,R}$ and $d_{i, R}$,
which transform as $-1$. The role of this $Z_2$ symmetry is to prevent
the Higgs doublets $\Phi_i^d$ from coupling the up-type quarks and
neutrino sector, and the $\Phi^u_i$ Higgs doublets from the down-type
quarks and charged leptons.

In addition we need scalar singlets, for example the $\chi_i$,
$i = 1, 2, 3$. These are gauge singlets transforming as a triplet
under the $A_4$ and trivially under $Z_4$. We also add two other gauge
singlet scalars $\zeta$ and $\eta$ both of which transform trivially
under $A_4$ but carry $Z_4$ charges $\mathbf{z}$ and $\mathbf{z}^2$
respectively.
Notice that, since under the $Z_4$ symmetry the field $\eta$ carries a
charge $\mathbf{z}^2 = -1$, it follows that $\eta$ can be taken to be
real.

As discussed in \cite{Chulia:2016ngi, Chulia:2016giq} the lepton
quarticity symmetry $Z_4$ serves a double purpose. It not only ensures
that neutrinos are Dirac particles, but also guarantees the stability
of the scalar particle $\zeta$, making it a viable dark matter
WIMP. If the quarticity symmetry is broken either by an explicit soft
term or spontaneously, through non-zero vacuum expectation values
(vevs) to any of the scalars $\eta, \zeta$ which carry a non-trivial
$Z_4$ charge, then both the Dirac nature of neutrinos and stability of
dark matter is simultaneously lost.

We now turn our attention to the Yukawa sector of our model. In the
neutrino sector the Yukawa terms relevant for generating masses for
the neutrinos and the heavy neutral fermions $N_{L}, N_{R}$ are given
by
\begin{eqnarray}
\mathcal{L}_{\rm{Yuk}, \nu}  & = & y_1 \, \left [  \left( \begin{array}{c} \bar{L}_e  \\ \bar{L}_\mu \\ \bar{L}_\tau \end{array} \right)_{3} \, \otimes \,  \left( \begin{array}{c}  N_{1,R}   \\ N_{2, R} \\ N_{3, R} \end{array} \right)_{3}  \right]_1 \, \otimes \ \left ({\Phi}^u_1 \right)_1   
  \, + \, 
y_2 \,  \left [  \left( \begin{array}{c} \bar{L}_e  \\ \bar{L}_\mu \\ \bar{L}_\tau \end{array} \right)_{3} \, \otimes \,  \left( \begin{array}{c}  N_{1,R}   \\ N_{2, R} \\ N_{3, R} \end{array} \right)_{3}  \right]_{1''} \, \otimes \,  \left ({\Phi}^u_2 \right)_{1'}  
  \nonumber \\
& + &    y_3 \, \left [  \left( \begin{array}{c} \bar{L}_e  \\ \bar{L}_\mu \\ \bar{L}_\tau \end{array} \right)_{3} \, \otimes \,  \left( \begin{array}{c}  N_{1,R}   \\ N_{2, R} \\ N_{3, R} \end{array} \right)_{3}  \right]_{1'} \, \otimes \,   \left ({\Phi}^u_3 \right)_{1''}  
  \, + \,
y'_1 \,   \left [ \left( \begin{array}{c}  \bar{N}_{1,L}   \\ \bar{N}_{2, L} \\ \bar{N}_{3, L} \end{array} \right)_{3} \, \otimes \, \left( \begin{array}{c} \chi_1   \\ \chi_2 \\ \chi_3 \end{array} \right)_{3} \right]_1  \, \otimes \, \left(\nu_{e,R} \right)_1
  \nonumber \\
  & + &
y'_2 \,   \left [ \left( \begin{array}{c}  \bar{N}_{1,L}   \\ \bar{N}_{2, L} \\ \bar{N}_{3, L} \end{array} \right)_{3} \, \otimes \, \left( \begin{array}{c} \chi_1   \\ \chi_2 \\ \chi_3 \end{array} \right)_{3} \right]_{1''}  \, \otimes \, \left(\nu_{\mu,R} \right)_{1'}
   \, + \, 
y'_3 \,   \left [ \left( \begin{array}{c}  \bar{N}_{1,L}   \\ \bar{N}_{2, L} \\ \bar{N}_{3, L} \end{array} \right)_{3} \, \otimes \, \left( \begin{array}{c} \chi_1   \\ \chi_2 \\ \chi_3 \end{array} \right)_{3} \right]_{1'}  \, \otimes \, \left(\nu_{\tau,R} \right)_{1''}
    \nonumber \\
& + &     M \,   \left[ \left( \begin{array}{c}  \bar{N}_{1,L}   \\ \bar{N}_{2, L} \\ \bar{N}_{3, L} \end{array} \right)_{3} \,
\otimes \,\left( \begin{array}{c}  N_{1,R}   \\ N_{2, R} \\ N_{3, R} \end{array} \right)_{3} \right]_1 
      \, + \, 
c_1 \, \left[ \left( \begin{array}{c}  \bar{N}_{1,L}   \\ \bar{N}_{2, L} \\ \bar{N}_{3, L} \end{array} \right)_3  \, \otimes \, \left [  \left( \begin{array}{c}  \chi_{1}   \\ \chi_{2} \\ \chi_{3} \end{array} \right)_3 \, \otimes  \left( \begin{array}{c} N_{1,R}   \\ N_{2,R} \\ N_{3, R} \end{array} \right)_{3}   \right]_{3S} \right ]_1
    \nonumber \\
&+&    c_2 \, \left[ \left( \begin{array}{c}  \bar{N}_{1,L}   \\ \bar{N}_{2, L} \\ \bar{N}_{3, L} \end{array} \right)_3  \, \otimes \, \left [  \left( \begin{array}{c}  \chi_{1}   \\ \chi_{2} \\ \chi_{3} \end{array} \right)_3 \, \otimes  \left( \begin{array}{c} N_{1,R}   \\ N_{2,R} \\ N_{3, R} \end{array} \right)_{3}   \right]_{3A} \right ]_1 + h.c.
 \label{neutyuk}
\end{eqnarray}
where $3S$ and $3A$ denote the symmetric and antisymmetric $A_4$
triplet combinations obtained from the tensor product of two $A_4$
triplets. Notice also that $3S$ and $3A$ are not two different
irreducible representations of $A_4$, which only has one triplet, but
simply different contractions with the same transformation rule. Also,
$y_i, y'_i, c_1, c_2$; $i = 1, 2, 3$ are the Yukawa couplings which,
for simplicity, are taken to be real. The parameter $M$ is the gauge
and flavor-invariant mass term for the heavy leptons. Here we like to
highlight the important role played by the $A_4$ flavor
symmetry. Owing to the $A_4$ charges of the left and right handed
neutrinos, a tree level Yukawa coupling between them of type
$y_{\nu} \, \bar L_L \nu_R \Phi^u_i$ is forbidden. Thus neutrino
masses can only appear through type-I Dirac seesaw mechanism as we now
discuss.
 
After symmetry breaking the scalars $\chi_i$ and $\Phi^u_i$ acquire
vevs $\vev{ \chi_i} = u_i$; $\vev{ \Phi^u_i} = v_i^u$; $i = 1, 2, 3$.
 The invariant mass term $M$ can be naturally much larger than the
 symmetry breaking scales, i.e.  $ M \gg v^u_i, u_i$. In this limit,
 for any numerical purpose the last two terms in Eq.~\ref{neutyuk} can
 be safely neglected. Under this approximation the $6\times 6$ mass
 matrix for the neutrinos and the heavy neutral fermions in the basis
 $( \bar{\nu}_{e,L}, \bar{\nu}_{\mu,L}, \bar{\nu}_{\tau,L},
 \bar{N}_{1,L}, \bar{N}_{2,L}, \bar{N}_{3,L} )$
 and
 $( \nu_{e,R}, \nu_{\mu,R}, \nu_{\tau,R}, N_{1,R}, N_{2,R}, N_{3,R}
 )^T$ is given by
  \begin{eqnarray}
   M_{\nu, N} \, \, = \, \left( 
\begin{array}{cccccc}
0           & 0                   &  0                  & a'_1	 	& 0            &   0      \\
0           & 0                   &  0                  & 0	        & a'_2	       &   0       \\
0           & 0                   &  0                  & 0	 	& 0	       &   a'_3    \\ 
y'_1 u_1    & y'_2 u_1            &  y'_3 u_1           & M             & 0	       &  0	    \\
y'_1 u_2    & \omega y'_2 u_2     & \omega^2 y'_3 u_2   & 0		& M            &  0	    \\
y'_1 u_3    & \omega^2 y'_2 u_3   & \omega y'_3 u_3     & 0		& 0	       &  M         \\
\end{array}
\right)
   \label{gneutmass}
  \end{eqnarray}
 where $\omega$ is the third root of unity, with $\omega^3 = 1$ and

   \begin{eqnarray}
   a'_1 & = & y_1 v_1^u  +  y_2 v_2^u  +  y_3 v_3^u \nonumber \\
   a'_2 &=&   y_1 v_1^u + \omega y_2 v_2^u + \omega^2y_3 v_3^u \nonumber \\
   a'_3 &=& y_1 v_1^u + \omega^2y_2 v_2^u +\omega y_3 v_3^u
  \end{eqnarray}

  As mentioned before, owing to the $A_4$ symmetry, a direct
    coupling between $\nu_L$ and $\nu_R$ is forbidden, leading to
    the vanishing of all entries in the upper left quadrant of
    Eq.~\ref{gneutmass}.  The mass matrix in Eq.~\ref{gneutmass} can
  be rewritten in a more compact form, as
  \begin{eqnarray}
   M_{\nu, N} \, \, = \, \left( 
\begin{array}{cc}
0     		     &\mathrm{diag}(a'_1, a'_2, a'_3)            \\
\mathrm{diag}(u_1, u_2, u_3) \sqrt{3}U_{\rm m}\mathrm{diag}(y'_1,y'_2,y'_3)    & M \mathrm{diag}(1,1,1)  \\
\end{array}
\right)
   \label{gneutmasscompact}
  \end{eqnarray}

  Where $U_{\rm m}$ is the usual magic matrix,
    \begin{eqnarray}
  U_{\rm m} = \dfrac{1}{\sqrt{3}} \, \left( 
\begin{array}{ccc}
1     & 1           &  1           \\
1     & \omega      &  \omega^2   \\
1     & \omega^2    &  \omega     \\  
\end{array}
\right)~.
   \label{magmat}
  \end{eqnarray}
  
\begin{figure}[!h]
 \centering
  \includegraphics[scale=0.4]{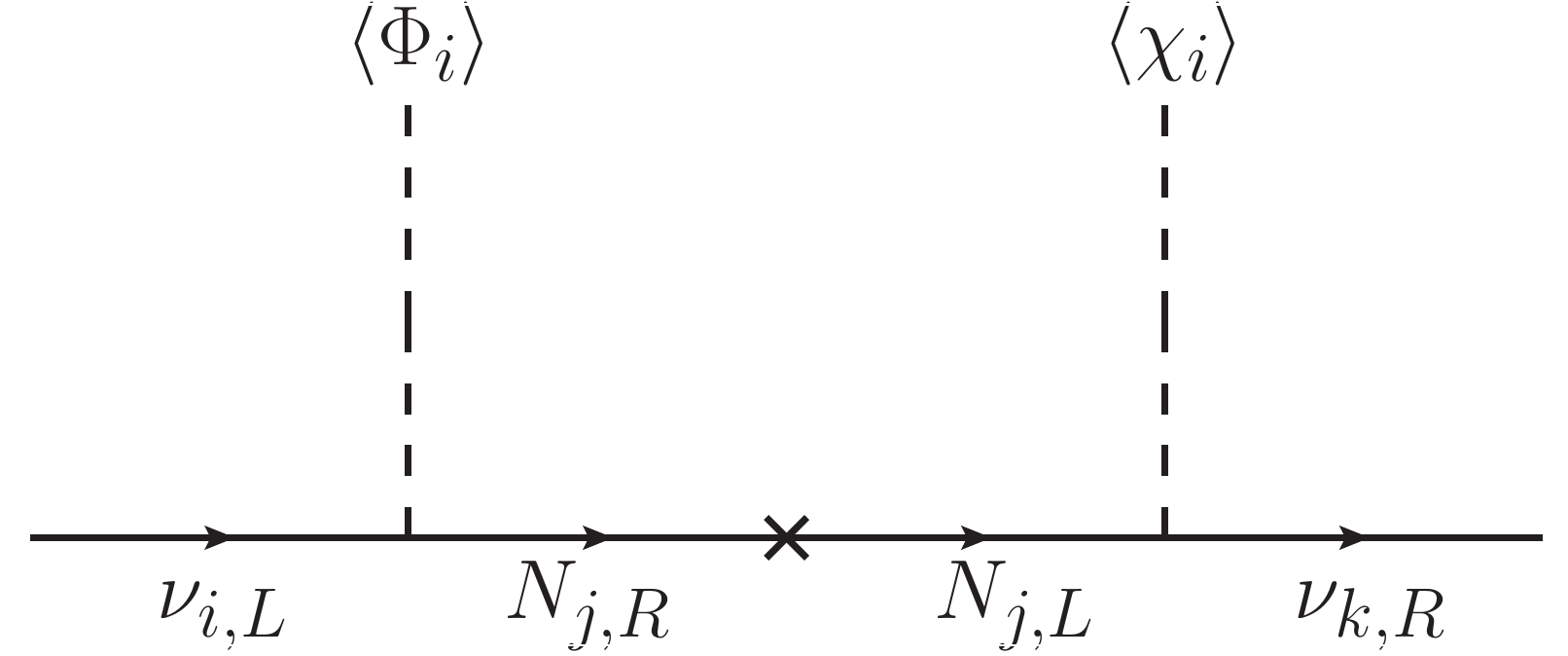}
  \caption{Feynman view of type-I Dirac seesaw mechanism in the model
    where the indices $i,j,k,m,l = 1,2,3$.} 
    \label{fig:feyn}
  \end{figure}

  Note that, in the limit $ M \gg v^u_i, u_i$ the mass matrix in
  Eq.~(\ref{gneutmass}) can be easily block diagonalized by the
  perturbative seesaw diagonalization method given in
  Ref.~\cite{Schechter:1981cv}. The resulting $3 \times 3$ mass matrix
  for light neutrinos can be viewed as the Dirac version of the well
  known type-I seesaw mechanism. The above mass generation mechanism
  can also be represented diagramatically as shown in
  Fig. \ref{fig:feyn}. 

  The $3\times 3$ matrix for the light neutrinos is
  \begin{eqnarray}
   M_{\nu} = \frac{1}{M}\mathrm{diag}(a_1, a_2, a_3)\sqrt{3}U_{\rm m}\mathrm{diag}(y'_1,y'_2,y'_3)~,
   \label{gnumass}
  \end{eqnarray}
  where $a_i = a'_i u_i$. We take the alignment $u_1 = u_2 = u_3 = u$
  for the vev of the $A_4$ triplet scalars $\chi_i$ similar
  to~\cite{Babu:2002dz, Ma:2001dn}. In this alignment limit of $A_4$
  triplet scalars we have 
   \begin{eqnarray}
   a_1 &=& (y_1 v_1^u + y_2v_2^u + y_3v_3^u) u \nonumber  \\
   a_2 &= &(y_1v_1^u + \omega y_2v_2^u + \omega^2y_3v_3^u) u \nonumber \\
   a_3 &=& (y_1v_1^u  + \omega ^2 y_2v_2^u + \omega y_3v_3^u ) u 
   \label{sima}
  \end{eqnarray}
  which simplifies the notation, although it does not change the form
  of the neutrino mass matrix in Eq.~(\ref{gnumass}). Notice that we
  have not imposed any alignment for the vevs of the $A_4$ singlet
  scalars $\Phi^u_i$~\footnote{Doing so for the different $A_4$
    singlet scalar vevs is not very natural. Indeed, unlike the case
    of $A_4$ triplet scalars, a priori the vevs of different $A_4$
    singlet scalars have no reason to obey any mutual alignment. }
  The light neutrino mass matrix of Eq.~\ref{gnumass} with the
  simplified $a_i$ of Eq.~\ref{sima} can be diagonalized by a
  bi-unitary transformation
\begin{eqnarray}
U_\nu^\dagger M_\nu V_\nu = D_\nu,
\label{biunit}
\end{eqnarray}
where $D_\nu$ is diagonal, real and positive. Owing to the $A_4$
flavor symmetry, the resulting rotation matrix acting on left handed
neutrinos $U_\nu$ in the standard parametrization (for both
hierarchies), leads to $\theta^\nu_{23} = \frac{\pi}{4}$ and
$\delta^\nu = \pm \frac{\pi}{2}$ while the other two angles can be
arbitrary. Thus, owing to the $A_4$ symmetry, $U_\nu$ in standard
parameterization leads to following mixing angles
\begin{eqnarray}
 \theta^\nu_{23} & = & 45^\circ,  \quad \quad \delta^\nu \, = \, \pm \, 90^\circ  \nonumber \\
 \theta^\nu_{12} & = & \rm{arbitrary} \quad \quad  \theta^\nu_{13} \, = \, \rm{arbitrary} 
 \label{unu}
\end{eqnarray}
Similar features of maximal $\theta_{23}$ and $\delta$ have been
obtained previously in the context of Majorana neutrinos
\cite{Babu:2002dz,Ma:2015pma}.  Although the angles $\theta^\nu_{12}$
and $\theta^\nu_{13}$ can take arbitrary values they are strongly
correlated with each other. 

We have performed an extensive numerical scan for both type of
hierarchies in the whole parameter range taking all Yukawa couplings
in the perturbative range of $[-1, 1]$. We find that in the whole
allowed range for either type of hierarchy, one cannot simultaneously
fit both $\theta^\nu_{12}$ and $\theta^\nu_{13}$ in the current global
experimental range obtained from neutrino oscillation
experiments~\cite{Forero:2014bxa}. This implies that in our model
$U_\nu$ alone cannot explain the current neutrino oscillation data.

However, the lepton mixing matrix $U_{LM}$ which is probed by neutrino
oscillation experiments is the product of the charged lepton rotation
matrix $U_l$ with the neutrino transformation matrix
$U_\nu$~\cite{Schechter:1980gr} i.e.
\begin{eqnarray}
 U_{LM} & = & U^\dagger_l \, U_\nu
 \label{ulm}
\end{eqnarray}
In our model the charged lepton mixing matrix $U_l$ is also
non-trivial and contributes to the full leptonic mixing matrix
$U_{LM}$. We now move to discuss the structure of mass matrices and
mixing matrices for charged leptons as well as the up and down type
quarks.  \\[-.2cm]

We now turn to the discussion with up type quark mass matrix.  The
invariant Yukawa Lagrangian relevant to generating up type quark mass
matrix is given by
   \begin{eqnarray}
  \mathcal{L}_{\rm{Yuk}, u}  & = & y^u_1 \, \left [  \left( \begin{array}{c} \bar{Q}_1  \\ \bar{Q}_2 \\ \bar{Q}_3 \end{array} \right)_{3} \, \otimes \,  \left( \begin{array}{c}  u_{R}   \\ c_{R} \\ t_{R} \end{array} \right)_{3}  \right]_1 \, \otimes \ \left ({\Phi}^u_1 \right)_1   
  \, + \,    y^u_2 \,  \left [  \left( \begin{array}{c} \bar{Q}_1  \\ \bar{Q}_2 \\ \bar{Q}_3 \end{array} \right)_{3} \, \otimes \,  \left( \begin{array}{c}  u_{R}   \\ c_{R} \\ t_{R} \end{array} \right)_{3}  \right]_{1''} \, \otimes \,  \left ({\Phi}^u_2 \right)_{1'} \nonumber \\
  & + & y_3^u \, \left [  \left( \begin{array}{c} \bar{Q}_1  \\ \bar{Q}_2 \\ \bar{Q}_3 \end{array} \right)_{3} \, \otimes \,  \left( \begin{array}{c}  u_{R}   \\ c_{R} \\ t_{R} \end{array} \right)_{3}  \right]_{1'} \, \otimes \,   \left ({\Phi}^u_3 \right)_{1''}   + \, \, h.c. 
  \label{uquarkyuk}
 \end{eqnarray}
 where $y^u_i$; $i = 1,2,3$ are the Yukawa couplings which for
 simplicity we take to be all real.  After spontaneous symmetry
 breaking Eq.~\ref{uquarkyuk} leads to a diagonal mass matrix given by
\begin{eqnarray}
 M_u & = & \left( 
\begin{array}{ccc}
y^u_1 v^u_1 + y^u_2 v^u_2  + y^u_3 v^u_3   & 0               &   0       \\
0            & y^u_1 v^u_1 + \omega \, y^u_2 v^u_2  + \omega^2 \, y^u_3 v^u_3         &   0     \\
0            & 0           &   y^u_1 v^u_1 + \omega^2 \, y^u_2 v^u_2  + \omega \, y^u_3 v^u_3           \\  
\end{array}
\right)~.
 \label{upmassmat}
\end{eqnarray}

On the other hand, the Yukawa Lagrangian relevant to down type quarks
mass generation is given by
  \begin{eqnarray}
  \mathcal{L}_{\rm{Yuk}, d}  & = & y^d_1 \, \left[ \left( \begin{array}{c}  \bar{Q}_{1}   \\ \bar{Q}_{2} \\ \bar{Q}_{3} \end{array} \right)_3  \, \otimes \, \left [  \left( \begin{array}{c}  q_{d,R}   \\ q_{s, R} \\ q_{b, R} \end{array} \right)_3 \, \otimes  \left( \begin{array}{c} \Phi_1^d   \\ \Phi_2^d \\ \Phi_3^d \end{array} \right)_{3}   \right]_{3S} \right ]_1
    \nonumber \\
  &+&    y^d_2 \, \left[ \left( \begin{array}{c}  \bar{Q}_{1}   \\ \bar{Q}_{2} \\ \bar{Q}_{3} \end{array} \right)_3  \, \otimes \, \left [  \left( \begin{array}{c} q_{d,R}   \\ q_{s, R} \\ q_{b, R} \end{array} \right)_3 \, \otimes  \left( \begin{array}{c} \Phi_1^d   \\ \Phi_2^d \\ \Phi_3^d \end{array} \right)_{3}   \right]_{3A} \right ]_1 + h.c.
    \label{quarkyuk}
 \end{eqnarray}
 where $y^d_i$; $i = 1,2$ are the Yukawa couplings which for
 simplicity are taken to be real. The resulting mass matrix for down
 type quarks after spontaneous symmetry breaking is given by
  \begin{eqnarray}
   M_{d} & = & \left( 
\begin{array}{ccc}
0              & a_d \alpha      &   b_d       \\
b_d \alpha     & 0               &   a_d r     \\
a_d            & b_d r           &   0          \\  
\end{array}
\right)~.
   \label{gquarkmass}
  \end{eqnarray}
  \noindent
  where $\vev{ \Phi^d_i} = v^d_i$; $i = 1,2,3$ and
  $a_d = (y_1^d-y_2^d) v_2^d$, $b_l = (y_1^d +y_2^d) v_2^d$. Moreover,
  $\alpha$ and $r$ are ratios of the vevs of $\Phi^d_i$ and are given
  as $\alpha = \rfrac{v_3^d}{v_2^d}$ and $r = \rfrac{v_1^d}{v_2^d}$.

Finally, the invariant Yukawa terms for the charged leptons is given by
 \begin{eqnarray}
  \mathcal{L}_{\rm{Yuk}, l}  & = & y^l_1 \, \left[ \left( \begin{array}{c}  \bar{L}_{1}   \\ \bar{L}_{2} \\ \bar{L}_{3} \end{array} \right)_3  \, \otimes \, \left [  \left( \begin{array}{c}  l_{e,R}   \\ l_{\mu, R} \\ l_{\tau, R} \end{array} \right)_3 \, \otimes  \left( \begin{array}{c} \Phi_1^d   \\ \Phi_2^d \\ \Phi_3^d \end{array} \right)_{3}   \right]_{3S} \right ]_1
    \nonumber \\
  &+&    y^l_2 \, \left[ \left( \begin{array}{c}  \bar{L}_{1}   \\ \bar{L}_{2} \\ \bar{L}_{3} \end{array} \right)_3  \, \otimes \, \left [  \left( \begin{array}{c}  l_{e,R}   \\ l_{\mu, R} \\ l_{\tau, R} \end{array} \right)_3 \, \otimes  \left( \begin{array}{c} \Phi_1^d   \\ \Phi_2^d \\ \Phi_3^d \end{array} \right)_{3}   \right]_{3A} \right ]_1 + h.c.
    \label{lepyuk}
 \end{eqnarray}
 where $y^l_i$, $i = 1, 2$, are the Yukawa couplings which, for
 simplicity, we take to be real. After symmetry breaking the charged
 lepton mass matrix is given by
  \begin{eqnarray}
   M_{l} = \left( 
\begin{array}{ccc}
0               & a_l \alpha           &   b_l \\
b_l \alpha      & 0                    &   a_l r \\
a_l             & b_l r                &   0  \\  
\end{array}
\right)~.
   \label{glepmass}
  \end{eqnarray}
  where, just as in the down quark case, here also
  $a_l = (y_1^l-y_2^l) v_2^d$, $b_l = (y_1^l +y_2^l) v_2^d$.  
  The parameters $\alpha, r$ which are the ratios of the vevs of
  $\Phi^d_i$ i.e.  $\alpha = \rfrac{v_3^d}{v_2^d}$ and
  $r = \rfrac{v_1^d}{v_2^d}$ are the same as those defined after
  Eq.~\ref{gquarkmass}. This matrix is completely analogous to the
  down-type quark mass matrix. Note that while $\alpha$ and $r$ are the same
  both in the quark and in the lepton sector, as they are simply
  ratios between the vevs of $\Phi_i^d$, while $a_f$ and $b_f$,
  $f \in \{l, q\}$, are different.

  These mass matrices for charged leptons and down-type quarks
  correspond to those discussed
  in~\cite{Morisi:2011pt,King:2013hj,Morisi:2013eca,Bonilla:2014xla}
  and lead to the generalized bottom-tau relation of \ref{eq:b-tau}.
  In section \ref{sec:numerical-scan} we show that there is enough
  freedom to fit the charged lepton and down type quark masses within
  their 1-$\sigma$ range. Apart from fitting all the masses as well as
  leading to the generalized bottom-tau relations, the charged lepton
  mass matrix \ref{glepmass} also leads to non-trivial charged lepton
  rotation matrix $U_l$.  As we show in section
  \ref{sec:numerical-scan} this non-trivial contribution from $U_l$
  results in a lepton mixing matrix $U_{LM}$ consistent with the
  current global fits to neutrino oscillation data
  \cite{Forero:2014bxa}. The lepton mixing
  matrix obtained from our model also implies normal hierarchy for
  neutrino masses and leads to an interesting correlation between the
  atmospheric mixing angle and CP violating phase, the two most
  ill--determined parameters in leptonic mixing matrix.
  
 \section{Flavor predictions: numerical results}
\label{sec:numerical-scan} 

In this section we discuss the phenomenological implications of our
model.  The important predictions emerging in our model are: a) the
flavor-dependent bottom-tau unification mass relation of
Eq.~(\ref{eq:b-tau}) b) a correlation between the two poorly
determined oscillation parameters: the atmospheric angle $\theta_{23}$
and $\delta_{CP}$ and c) a normal hierarchy for the neutrinos.
In this section we discuss these numerical predictions in some detail, 
given the experimentally measured ``down-type'' fermion masses, solar and
reactor mixing angles as well as neutrino squared mass differences.

\subsection{Charged lepton and down-type quark masses}
\label{clmass}

We start our discussion by looking in more detail at the down type
quark and charged lepton mass matrices discussed previously in
Eqs.~\ref{gquarkmass} and \ref{glepmass}. This structure for the down
type quark and charged lepton mass matrices has been previously
discussed in several works
\cite{Morisi:2011pt,King:2013hj,Morisi:2013eca,Bonilla:2014xla}. In
this section for illustration purpose we first discuss the results
obtained in previous works by closely following the approach taken in
previous works like in \cite{Morisi:2009sc}. Subsequently, we will
generalize the analysis of previous works and discuss how the same
results can be obtained using a more general setup and more detailed
considerations.

We start from the charged lepton mass matrix obtained in
Eq.~\ref{glepmass}. The correct charged lepton masses are reproduced
if the vevs of the $A_4$ triplet fields $\Phi^d_i$ satisfy the
alignment limit $v^d(1, \epsilon_1, \epsilon_2)$, where
$v^d \gg \epsilon_1, \epsilon_2$. Then, in similar notation and spirit
as in Ref.~\cite{Morisi:2009sc}, we extract the three invariants of
the Hermitian matrix $S = M_l M_l^\dagger$: $Det(S)$, $Tr(S)$ and
$Tr(S)^2-Tr(S^2)$. We then compute their values in the diagonal
basis in terms of the charged lepton masses, $m_e$, $m_\mu$ and
$m_\tau$. The equations are
   \begin{eqnarray}
   \textnormal{Det} S &=& (m_e m_\mu m_\tau)^2 \nonumber \\
   \textnormal{Tr} S& =& m_e^2 + m_\mu^2 + m_\tau^2 \nonumber  \\
    (\textnormal{Tr} S)^2 - \textnormal{Tr} S^2 &=& 2m_e^2m_\mu^2 + 2m_e^2 m_\tau^2 + 2m_\mu^2 m_\tau^2
    \label{masseq}
  \end{eqnarray}
 
  The expressions in~Eq.\ref{masseq} can be readily solved in the
  vev alignment limit $v^d(1, \epsilon_1, \epsilon_2)$ discussed
  before.
  This amounts to the approximation 
  $$v_1^d \gg v_3^d~~~~~ {\rm and}~~~~~\dfrac{v_1^d}{v_2^d} \gg \dfrac{y_1^l+y_2^l}{y_1^l-y_2^l} \gg 1,$$
or equivalently  $$r \gg \alpha~~~~~{\rm and}~~~~~
  r \gg \dfrac{b_l}{a_l} \gg 1.$$
  The solutions for $r$, $a_l$ and $b_l$ are given as
\begin{eqnarray}
   r & \,\,=\,\, & \frac{m_\tau}{\sqrt{m_e m_\mu}} \sqrt{\alpha}  \label{rleq} \\
   a_l & \,\,=\,\, & \frac{m_\mu}{m_\tau} \sqrt{\frac{m_e m_\mu}{\alpha}}  \label{aleq} \\
    b_l & \,\,=\,\, & \sqrt{\frac{m_e m_\mu}{\alpha}}  \label{bleq}
\end{eqnarray}

Owing to the $A_4$ symmetry, the charged lepton mass matrix in
Eq.~\ref{glepmass} and down type quark mass matrix in
Eq.~\ref{gquarkmass} have the same structure. As a result the down
quark mass matrix can also be decomposed using equations analogous to
Eq.~\ref{masseq}. For down type mass matrix of \ref{gquarkmass} we
obtain
\begin{eqnarray}
r & = & \frac{m_b}{\sqrt{m_d m_s}} \sqrt{\alpha}  \label{rdeq} \\
a_d & = & \frac{m_s}{m_b} \sqrt{\frac{m_d m_s}{\alpha}}  \label{adeq} \\
b_d & = & \sqrt{\frac{m_d m_s}{\alpha}}  \label{bdeq}
\end{eqnarray}
 Note that the parameters $\alpha$ and $r$ are common for both the
 charged lepton sector as well as in the down-type quark sector, as
 they are simply ratios between vevs of the fields $\Phi^d_i$. Thus
 comparing Eqs.~\ref{rleq} and \ref{rdeq} we obtain the following mass
 relation
   \begin{eqnarray} \label{massrelation}
\frac{m_\tau}{\sqrt{m_e m_\mu}} & =  & \frac{m_b}{\sqrt{m_s m_d}}
  \end{eqnarray}

  The procedure sketched above can be performed in a more general way
  by solving the equations numerically. The relevant equations for the
  case of charged leptons are
\begin{eqnarray} \label{massequations}
(m_e m_\mu m_\tau)^2 & = & {a_l}^6r^2 \alpha^2 + 2 {a_l}^3 b_l^3 r^2 \alpha^2+b_l^6 r^2 \alpha^2 
  \nonumber \\ 
m_e^2 + m_\mu^2 + m_\tau^2 & = & ({a_l}^2 + b_l^2)(1+r^2+\alpha^2)  \nonumber \\
 2m_e^2m_\mu^2 + 2m_e^2 m_\tau^2 + 2m_\mu^2 m_\tau^2  & = &
({a_l}^2 + b_l^2)^2(1+r^2+\alpha^2)^2  -({a_l}^2 +b_l^2r^2)^2  - (b_l^2+{a_l}^2 \alpha^2)^2
 - ({a_l}^2 r^2+ b_l^2 \alpha^2)^2 
\end{eqnarray}
Taking as input parameters the best fit values (at $M_Z$ scale) for
the charged lepton masses~\cite{Xing:2011aa} and imposing
$r > \frac{b}{a}$, there is a one-parameter family of solutions to
these equations.  These are related to the approximate solution
described before.
  We build the functions $r(\alpha)$, $a_l(\alpha)$ and $b_l(\alpha)$
  taking $\alpha$ as a free parameter.  
  In the correct range for the parameter $\alpha$, the unique solution
  is found to be near the limit $r \gg \dfrac{b}{a} \gg 1$ and
  therefore it again leads to the mass relation in
  Eq.~\ref{eq:b-tau}. Since the $(\alpha, r, a_l, b_l)$ are solutions
  of Eqs.~\ref{massequations}, the charged lepton masses are fitted
  exactly to their best-fit values.
  In order to underpin the relevant solution for down type quark
  masses, we also need to take into account not only the mass relation
  \ref{eq:b-tau} and the charged lepton masses, but also the
  constraints for the experimental measurements (along with
  renormalization group evolution to $M_Z$ scale) of all the down-type
  quark masses \cite{Xing:2011aa}. Here, we will impose the rather
  stringent 1-$\sigma$ bounds\footnote{Imposing 1-$\sigma$ is in
    fact rather stringent, and can easily be relaxed to a more
    conservative criterium e.g. 3-$\sigma$. We have deliberatively
    imposed the stringent 1-$\sigma$ bound in other to highlight the
    high precision obtained from our results. } on the down-type
quarks masses at $Z$ boson energy scale \cite{Xing:2011aa}.
    
Then, for each valid $( \alpha, r, a_l, b_l,)$ we take $a_q$ and $b_q$ as 
   \begin{eqnarray}
   a_q = \frac{m_s}{m_b} \sqrt{\frac{m_d m_s}{\alpha}} (1+\epsilon_1) \\
   b_q = \sqrt{\frac{m_d m_s}{\alpha}}(1+\epsilon_2)~,
  \end{eqnarray}
  where $\epsilon_1$ and $\epsilon_2$ are expected to be small.
  
\begin{figure}[!h]
 \centering
  \includegraphics[scale=0.75]{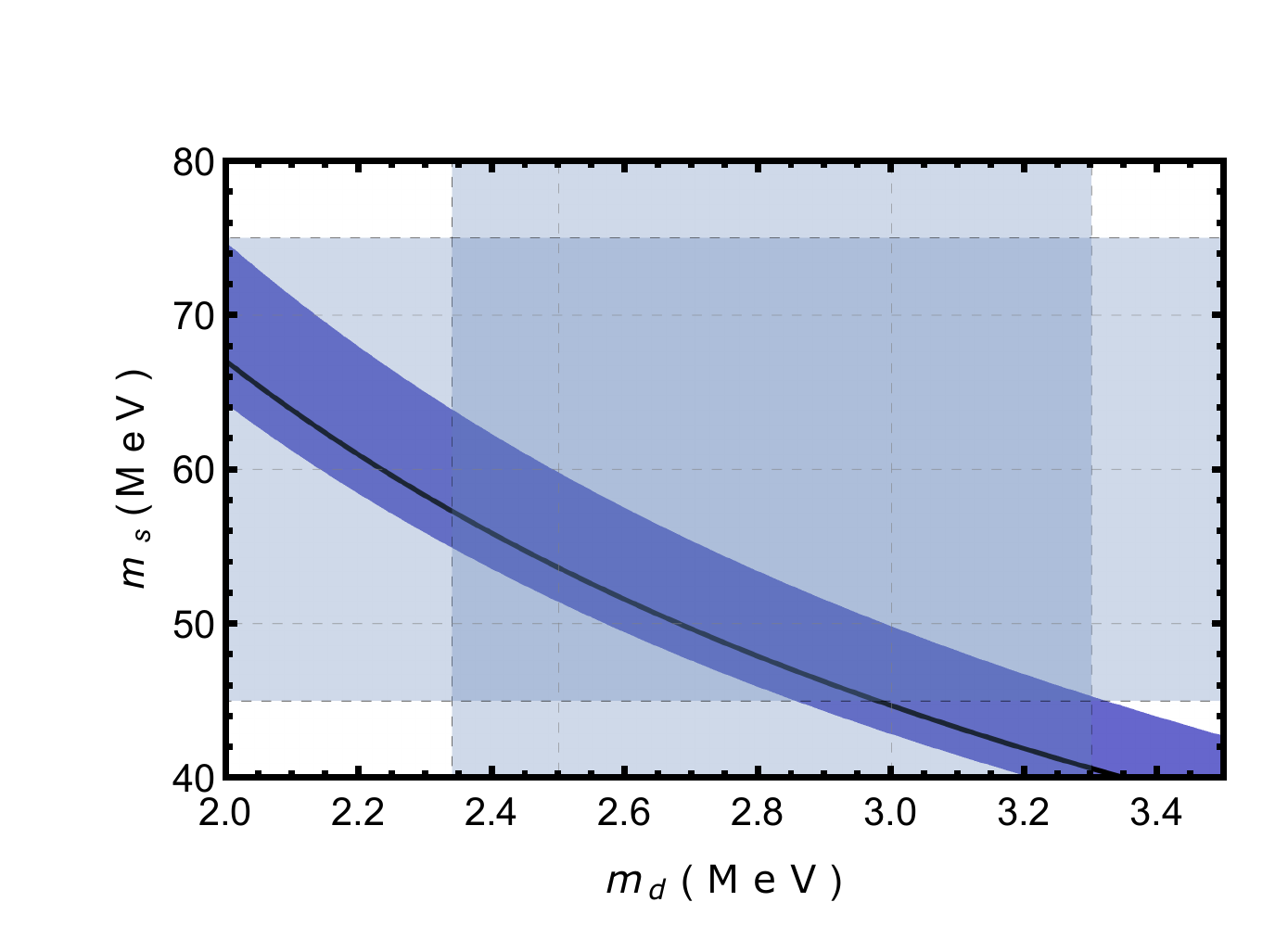}
  \caption{Prediction for the $s$ and $d$ quark masses (at $M_Z$
    scale) in our model. The dark blue area is the allowed region from
    our model for the $s$ and $d$ quark masses, while varying the mass
    of the $b$ quark in its 1-$\sigma$ range. The light blue area is
    the allowed 1-$\sigma$ range (at $M_Z$ scale) for the mass of the
    quarks $s$ and $d$ \cite{Xing:2011aa}.}
  \label{fig:msvsmd}
  \end{figure}

  Using this procedure gives sets of parameters
  $(\alpha, r, a_l, b_l, a_q, b_q)$ which give at the same time best
  fit values for charged lepton masses, down-type quarks inside the
  $1\sigma$ range and the mass relation in Eq.~\ref{eq:b-tau}.  In
  figure \ref{fig:msvsmd} we show the family-dependent bottom-tau mass
  prediction of our model for the s and d masses, along with their
  allowed 1-$\sigma$ ranges.

\subsection{The charged piece of th lepton mixing matrix $U_l$ }
\label{cul}

The charged lepton mass matrix Eq.~\ref{glepmass} not only leads to
correct lepton masses but also to non-trivial charged lepton rotation
matrix $U_l$ as we discuss now. Just like the neutrino mass matrix,
the charged lepton mass matrix can also be diagonalized by a
bi-unitary transformation as
\begin{eqnarray}
U_l^\dagger M_l V =\mathrm{diag}(m_e, m_\mu, m_\tau) 
\end{eqnarray}

The charged lepton mixing matrix $U_l$ in standard
parameterization can be written as
\begin{eqnarray}
 U_l = P U_{23} (\theta^l_{23}, 0) U_{13} (\theta^l_{23}, \delta^l) U_{12} (\theta^l_{12}, 0) P'
\end{eqnarray}
where $P$ and $P'$ are diagonal matrix of phases and $U_{ij}$ is the
usual complex rotation matrix appearing in the symmetrical
parametrization of fermion mixing given in~\cite{Schechter:1980gr}, e.g.
\begin{eqnarray}
  U_{12} (\theta_{12}, \delta) =  \, \left( 
\begin{array}{ccc}
\cos \theta_{12}      & e^{-i\delta} \sin \theta_{12}            &  0           \\
-e^{i\delta}\sin \theta_{12}     & \cos \theta_{12} &  0   \\
0     & 0    &  1     \\  
\end{array}
\right)~.
   \label{magmat}
  \end{eqnarray}
  with an analogous definitions for $U_{13}$ and $U_{23}$.  For the
  charged lepton mass matrix \ref{glepmass}, we find that
\begin{eqnarray}
\sin \theta^l_{12} & = & \sqrt{\frac{m_e}{m_\mu}} \frac{1}{\sqrt{\alpha}}+ \mathcal{O}(\frac{1}{\alpha^2}) \approx \mathcal{O} (\lambda_C) \nonumber \\
\sin \theta^l_{13} & = & \frac{m_u}{m_\tau^2} \sqrt{m_e m_\mu} \frac{1}{\sqrt{\alpha}} + \mathcal{O}(\frac{1}{\alpha^2}) \approx \mathcal{O} (10^{-5}) \nonumber \\
\sin \theta^l_{23} & = & \frac{m_e m_\mu^2}{m^3_\tau} \frac{1}{\alpha} + \mathcal{O}(\frac{1}{\alpha^2}) \approx \mathcal{O} (10^{-7})
 \label{leprot}
\end{eqnarray}
Where $\lambda_C \approx 0.22$ is the sine of the Cabbibo angle. In order to reproduce
adequate values for the CKM matrix elements we may introduce a
vector-like quark mixing with the up-type quarks, along the lines
followed recently in \cite{Bonilla:2016sgx}.

The diagonal phases in P and P' are all are exactly $0$ except for one
which is $\pi$. Performing the numerical computation reconfirms the
results obtained in Eq.~\ref{leprot} for the charged lepton mass
matrix i.e. $\theta^l_{12}$ is finite and its value depends on the
value of $\alpha$ in an inverse way, while $\theta^l_{13}$ and
$\theta^l_{23}$ are both negligible (in particular,
$\theta^l_{13} \sim 10^{-5}$ and $\theta^l_{23} \sim 10^{-7}$.  Then,
the charged lepton mixing matrix for our model is given as
\begin{eqnarray}
U_l \approx \left(\begin{matrix}
\cos \theta^l_{12} & \sin {\theta^l_{12}} & 0 \\
-\sin {\theta^l_{12}} & \cos \theta^l_{12} & 0 \\
0 & 0 & 1
\end{matrix} \right) \cdot 
\left(\begin{matrix}
-1 & 0 & 0 \\
0 & 1 & 0 \\
0 & 0 & 1
\end{matrix} \right) 
\end{eqnarray}

Thus in our model the lepton mixing matrix
$U_{LM} = U^\dagger_l U_\nu$ receives significant charged lepton
corrections which have interesting phenomenological consequences as we
discuss in next section.

\subsection{The lepton mixing matrix and  neutrino mass ordering}
\label{lm}

As mentioned before in Section \ref{sec:model}, the light neutrino
mass matrix in Eq.~\ref{gnumass} leads to the neutrino mixing matrix
$U_\nu$ which in standard parameterization~\cite{Schechter:1980gr}
leads to
\begin{eqnarray}
U_\nu  =  P U_{23}\left(\rfrac{\pi}{4}, 0\right)& U_{13}\left( \theta_{13}^\nu, \rfrac{\pi}{2}\right)& U_{12} \left( \theta_{12}^\nu, 0\right) P'
\label{rnu}
\end{eqnarray}
As mentioned before, owing to the $A_4$ symmetry, we have that
$\theta_{23}^\nu = 45^{\circ}$ and $\delta^\nu_{CP} = 90^{\circ}$ for
both types of mass ordering: normal hierarchy (NH) or inverted
hierarchy (IH).  
Since neutrinos in our model are Dirac fermions, the phases in the
right in Eq.~\ref{rnu} i.e.  $P'$, are unphysical, while
$\theta_{13}^\nu$ and $\theta_{12}^\nu$ are strongly correlated
between each other. 
This result is completely general and follows from the $A_4$ symmetry,
independently of the mass hierarchy, NH or IH. However, the behavior
of the correlation between $\theta_{12}^\nu$ and $\theta_{13}^\nu$
does depend on the choice of NH or IH.

Taking into account the results in the previous sections, the lepton
mixing matrix is
\begin{eqnarray}
U_{LM} = U_l^\dagger U_\nu = U_{12}\left(\theta^l_{12}, 0\right)^\dagger  \cdot U_{\nu}
\end{eqnarray}

\begin{figure}[!h]
 \centering
  \includegraphics[scale=0.505]{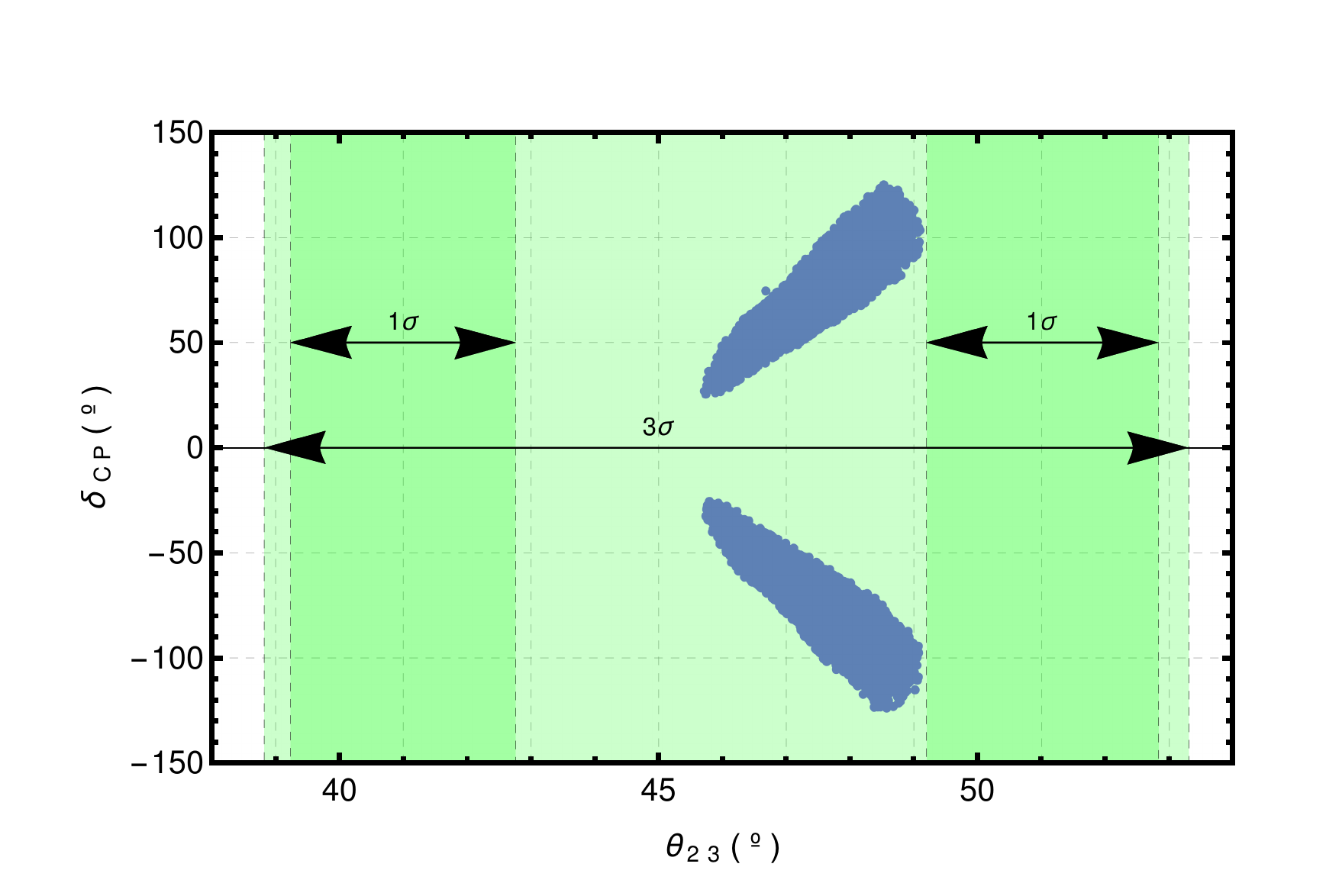}
  \includegraphics[scale=0.48]{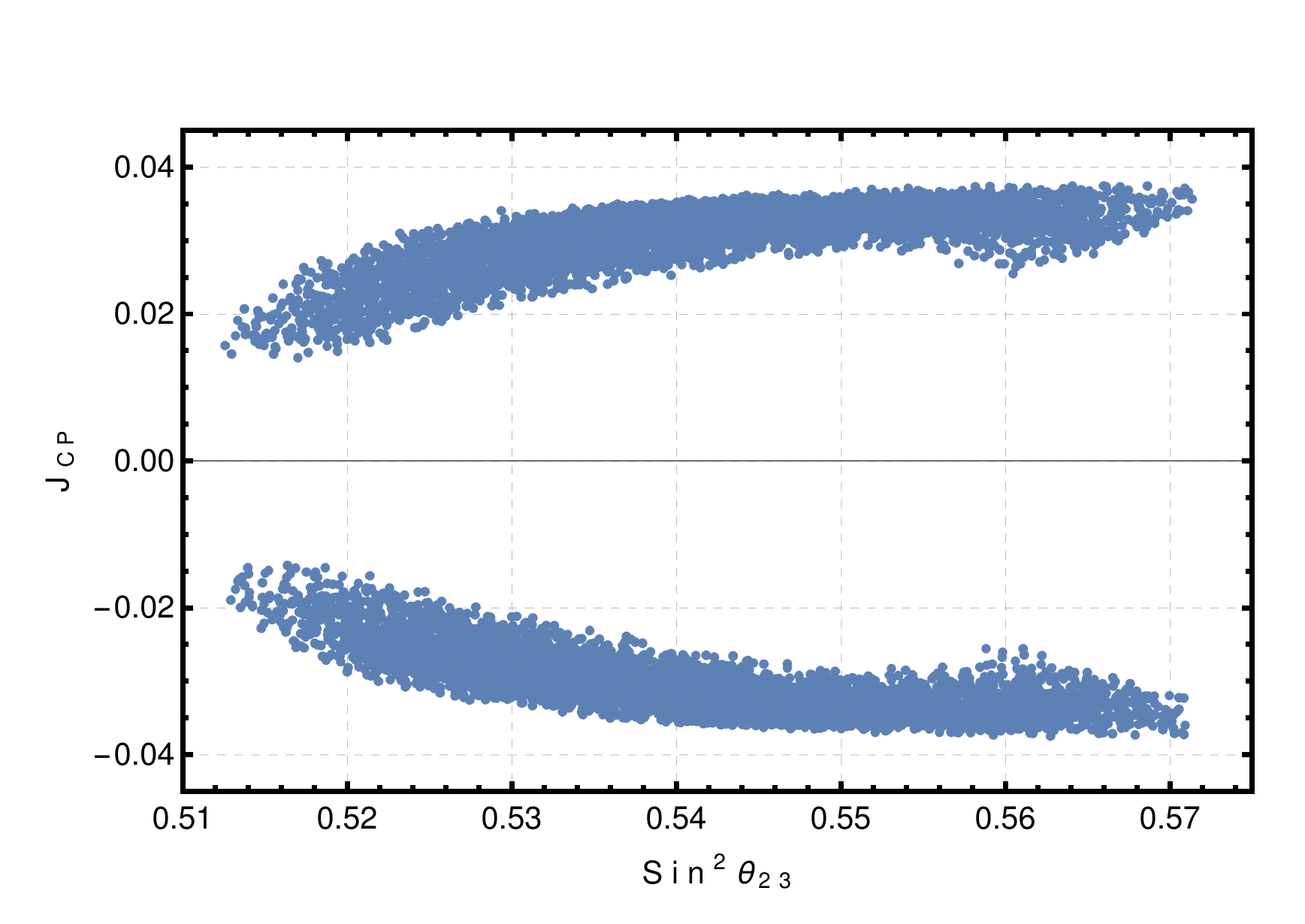}
  \caption{CP violation and $\theta_{23}$ predictions within the
    model. Left panel: $\delta_{CP}$ vs $\theta_{23}$. The green
    regions are the $1\sigma$ (dark) and $3\sigma$ (light) regions for
    $\theta_{23}$ from current oscillation fit. Right panel: Same
    correlation, now showing $J_{CP}$ vs $\sin^2 \theta_{23}$ and
    zooming in the region allowed by the model, fully consistent in
    the $2\sigma$ experimental range.}
    \label{fig:theta23vsdelta}
  \end{figure}

One can regard the matrix $U_l$ as a correction to the neutrino
mixing parameters obtained just by diagonalizing the neutrino mass
matrix.
For the NH case, the angle $\theta_{12}^l$ has to be big enough
($\sim >15 ^{\circ}$) so as to account for the correct mixing angles
of the lepton mixing matrix, but at the same time it has to remain
controlled ($< 20^{\circ}$) otherwise the down-type quark masses
cannot be fitted. This means that the parameter $\alpha$ has to be
between $0.04$ and $0.08$.
This lepton mixing matrix can fit the neutrino oscillation parameters
within $3 \sigma$ at the same time as the mass matrices fit the
down-type quarks and the neutrino squared mass differences in the
$1\sigma$ range and the charged lepton masses exactly. Once the lepton
mixing matrix is written in the standard parametrization, two
interesting features arise. On the one hand, $\theta_{23}>45^{\circ}$
and, on the other, a strong correlation appears between the
atmospheric angle $\theta_{23}$ and $\delta_{CP}$, as shown in figure
\ref{fig:theta23vsdelta}.
  
  For IH, a different scenario arises. As in the case for NH, lepton
  corrections cannot be very big otherwise the down-type quark masses
  will not be fitted. However, the structure of the correlation
  between $\theta_{12}^\nu$ and $\theta_{13}^\nu$ implies that for
  allowed charged lepton corrections, the reactor angle $\theta_{13}$
  is always outside the $3\sigma$ allowed range. Note that the model
  does not include any a priori theoretical bias in favour of normal
  hierarchy but it is a prediction of the model once one impose
  experimental constraints.

\section{Discussion and Summary }
\label{sec:summary-conclusions-}

We have proposed a $A_4 \otimes Z_4 \otimes Z_2$ flavor extension of
the \sm with naturally small Dirac neutrino masses. Our lepton
quarticity symmetry simultaneously forbids Majorana mass terms and
provides dark matter stability. The flavor symmetry plays a multiple
role, providing : 
(i) a generalized family-dependent bottom-tau mass relation, Eq.~(\ref{eq:b-tau}) and Fig.~\ref{fig:msvsmd},
(ii)  a natural realization of the type-I seesaw mechanism for Dirac neutrino masses, as the tree level Dirac Yukawa term between left and right handed neutrinos is forbidden,
(iii)  a very predictive flavor structure to the lepton mixing
matrix. The latter directly correlates the CP phase $\delta_{CP}$ and the atmospheric angle $\theta_{23}$, as shown in
Fig.~\ref{fig:theta23vsdelta}. This implies that 
(iv) CP must be significantly violated in neutrino oscillations, and the atmospheric angle $\theta_{23}$ lies in the second octant,
(v) only the normal neutrino mass ordering is realized.

Our approach provides an adequate pattern of neutrino mass and mixing
as well as a viable stable dark matter.
This is achieved while providing testable predictions concerning the
currently most relevant oscillation parameters, the atmospheric angle
$\theta_{23}$ and the CP phase $\delta_{CP}$, as well as a successful
family generalization of bottom-tau unification, despite the absence
of an underlying Grand Unified Theory.
Our lepton quarticity approach also leads to other interesting
phenomena such as neutrinoless quadruple beta decay ($0 \nu 4 \beta$),
which has recently been probed by the NEMO
collaboration~\cite{Arnold:2017bnh}. The intimate connection between
the Dirac nature of neutrinos and dark matter stability constitutes a
key feature of our model. Other phenomenological implications will be
taken up elsewhere.

\begin{acknowledgments}

  This research is supported by the Spanish grants FPA2014-58183-P,
  Multidark CSD2009-00064, SEV-2014-0398 (MINECO) and
  PROMETEOII/2014/084 (Generalitat Valenciana). The feynman diagram in
  Fig.\ref{fig:feyn} was drawn using Jaxodraw \cite{Binosi:2003yf}.

\end{acknowledgments}


\bibliographystyle{bib_style_T1}

\begin{thebibliography}{10}
\providecommand{\url}[1]{\texttt{#1}}
\providecommand{\urlprefix}{URL }
\providecommand{\eprint}[2][]{\url{#2}}

\bibitem{Morisi:2011pt}
S.~Morisi, E.~Peinado, Y.~Shimizu and J.~W.~F. Valle, \emph{{Relating quarks
  and leptons without grand-unification}}, Phys.Rev. \textbf{D84} (2011)
  036003, \MYhref[eprintLinks]{http://arxiv.org/abs/1104.1633}{{\ttfamily
  arXiv:1104.1633 [hep-ph]}}.

\bibitem{King:2013hj}
S.~King, S.~Morisi, E.~Peinado and J.~W.~F. Valle, \emph{{Quark-Lepton Mass
  Relation in a Realistic A4 Extension of the Standard Model}}, Phys. Lett. B
  \textbf{724} (2013) 68--72,
  \MYhref[eprintLinks]{http://arxiv.org/abs/1301.7065}{{\ttfamily
  arXiv:1301.7065 [hep-ph]}}.

\bibitem{Morisi:2013eca}
S.~Morisi et~al., \emph{{Quark-Lepton Mass Relation and CKM mixing in an A4
  Extension of the Minimal Supersymmetric Standard Model}},
  \MYhref[journalLinks]{http://dx.doi.org/10.1103/PhysRevD.88.036001}{Phys.Rev.
  }\MYhref[journalLinks]{http://dx.doi.org/10.1103/PhysRevD.88.036001}{\textbf{D88}
  (2013) 036001},
  \MYhref[eprintLinks]{http://arxiv.org/abs/1303.4394}{{\ttfamily
  arXiv:1303.4394 [hep-ph]}}.

\bibitem{Bonilla:2014xla}
C.~Bonilla, S.~Morisi, E.~Peinado and J.~W.~F. Valle, \emph{{Relating quarks
  and leptons with the $T_7$ flavour group}},
  \MYhref[journalLinks]{http://dx.doi.org/10.1016/j.physletb.2015.01.017}{Phys.
  Lett.
  }\MYhref[journalLinks]{http://dx.doi.org/10.1016/j.physletb.2015.01.017}{\textbf{B742}
  (2015) 99--106},
  \MYhref[eprintLinks]{http://arxiv.org/abs/1411.4883}{{\ttfamily
  arXiv:1411.4883 [hep-ph]}}.

\bibitem{Ma:2006km}
E.~Ma, \emph{{Verifiable radiative seesaw mechanism of neutrino mass and dark
  matter}},
  \MYhref[journalLinks]{http://dx.doi.org/10.1103/PhysRevD.73.077301}{Phys.Rev.
  }\MYhref[journalLinks]{http://dx.doi.org/10.1103/PhysRevD.73.077301}{\textbf{D73}
  (2006) 077301},
  \MYhref[eprintLinks]{http://arxiv.org/abs/hep-ph/0601225}{{\ttfamily
  arXiv:hep-ph/0601225 [hep-ph]}}.

\bibitem{Hirsch:2013ola}
M.~Hirsch et~al., \emph{{WIMP dark matter as radiative neutrino mass
  messenger}}, JHEP \textbf{1310} (2013) 149,
  \MYhref[eprintLinks]{http://arxiv.org/abs/1307.8134}{{\ttfamily
  arXiv:1307.8134 [hep-ph]}}.

\bibitem{Merle:2016scw}
A.~Merle et~al., \emph{{Consistency of WIMP Dark Matter as radiative neutrino
  mass messenger}},
  \MYhref[journalLinks]{http://dx.doi.org/10.1007/JHEP07(2016)013}{JHEP
  }\MYhref[journalLinks]{http://dx.doi.org/10.1007/JHEP07(2016)013}{\textbf{07}
  (2016) 013}, \MYhref[eprintLinks]{http://arxiv.org/abs/1603.05685}{{\ttfamily
  arXiv:1603.05685 [hep-ph]}}.

\bibitem{Hirsch:2012ym}
M.~Hirsch, D.~Meloni et~al., \emph{{Proceedings of the first workshop on Flavor
  Symmetries and consequences in Accelerators and Cosmology (FLASY2011)}}
  (2012), \MYhref[eprintLinks]{http://arxiv.org/abs/1201.5525}{{\ttfamily
  arXiv:1201.5525 [hep-ph]}}.

\bibitem{Morisi:2012fg}
S.~Morisi and J.~W.~F. Valle, \emph{{Neutrino masses and mixing: a flavour
  symmetry roadmap}}, Fortsch.Phys. \textbf{61} (2013) 466--492,
  \MYhref[eprintLinks]{http://arxiv.org/abs/1206.6678}{{\ttfamily
  arXiv:1206.6678 [hep-ph]}}.

\bibitem{King:2014nza}
S.~F. King et~al., \emph{{Neutrino Mass and Mixing: from Theory to
  Experiment}},
  \MYhref[journalLinks]{http://dx.doi.org/10.1088/1367-2630/16/4/045018}{New
  J.Phys.
  }\MYhref[journalLinks]{http://dx.doi.org/10.1088/1367-2630/16/4/045018}{\textbf{16}
  (2014) 045018},
  \MYhref[eprintLinks]{http://arxiv.org/abs/1402.4271}{{\ttfamily
  arXiv:1402.4271 [hep-ph]}}.

\bibitem{Ma:2001dn}
E.~Ma and G.~Rajasekaran, \emph{{Softly broken A(4) symmetry for nearly
  degenerate neutrino masses}}, Phys. Rev. \textbf{D64} (2001) 113012,
  \MYhref[eprintLinks]{http://arxiv.org/abs/hep-ph/0106291}{{\ttfamily
  hep-ph/0106291}}.

\bibitem{Babu:2002dz}
K.~S. Babu, E.~Ma and J.~W.~F. Valle, \emph{{Underlying A(4) symmetry for the
  neutrino mass matrix and the quark mixing matrix}},
  \MYhref[journalLinks]{http://dx.doi.org/10.1016/S0370-2693(02)03153-2}{Phys.
  Lett.
  }\MYhref[journalLinks]{http://dx.doi.org/10.1016/S0370-2693(02)03153-2}{\textbf{B552}
  (2003) 207--213},
  \MYhref[eprintLinks]{http://arxiv.org/abs/hep-ph/0206292}{{\ttfamily
  arXiv:hep-ph/0206292 [hep-ph]}}.

\bibitem{Morisi:2013qna}
S.~Morisi, D.~Forero, J.~C. Romao and J.~W.~F. Valle, \emph{{Neutrino mixing
  with revamped A4 flavour symmetry}}, Phys.Rev. \textbf{D88} (2013) 016003,
  \MYhref[eprintLinks]{http://arxiv.org/abs/1305.6774}{{\ttfamily
  arXiv:1305.6774 [hep-ph]}}.

\bibitem{Hirsch:2010ru}
M.~Hirsch, S.~Morisi, E.~Peinado and J.~Valle, \emph{{Discrete dark matter}},
  \MYhref[journalLinks]{http://dx.doi.org/10.1103/PhysRevD.82.116003}{Phys.Rev.
  }\MYhref[journalLinks]{http://dx.doi.org/10.1103/PhysRevD.82.116003}{\textbf{D82}
  (2010) 116003},
  \MYhref[eprintLinks]{http://arxiv.org/abs/1007.0871}{{\ttfamily
  arXiv:1007.0871 [hep-ph]}}.

\bibitem{Boucenna:2012qb}
M.~Boucenna et~al., \emph{{Predictive discrete dark matter model and neutrino
  oscillations}},
  \MYhref[journalLinks]{http://dx.doi.org/10.1103/PhysRevD.86.073008}{Phys.Rev.
  }\MYhref[journalLinks]{http://dx.doi.org/10.1103/PhysRevD.86.073008}{\textbf{D86}
  (2012) 073008},
  \MYhref[eprintLinks]{http://arxiv.org/abs/1204.4733}{{\ttfamily
  arXiv:1204.4733 [hep-ph]}}.

\bibitem{Chulia:2016ngi}
S.~{Centelles Chuli{\'a}}, E.~Ma, R.~Srivastava and J.~W.~F. Valle,
  \emph{{Dirac Neutrinos and Dark Matter Stability from Lepton Quarticity}},
  \MYhref[journalLinks]{http://dx.doi.org/10.1016/j.physletb.2017.01.070}{Phys.
  Lett.
  }\MYhref[journalLinks]{http://dx.doi.org/10.1016/j.physletb.2017.01.070}{\textbf{B767}
  (2017) 209--213},
  \MYhref[eprintLinks]{http://arxiv.org/abs/1606.04543}{{\ttfamily
  arXiv:1606.04543 [hep-ph]}}.

\bibitem{Chulia:2016giq}
S.~{Centelles Chuli{\'a}}, R.~Srivastava and J.~W.~F. Valle, \emph{{CP
  violation from flavor symmetry in a lepton quarticity dark matter model}},
  \MYhref[journalLinks]{http://dx.doi.org/10.1016/j.physletb.2016.08.028}{Phys.
  Lett.
  }\MYhref[journalLinks]{http://dx.doi.org/10.1016/j.physletb.2016.08.028}{\textbf{B761}
  (2016) 431--436},
  \MYhref[eprintLinks]{http://arxiv.org/abs/1606.06904}{{\ttfamily
  arXiv:1606.06904 [hep-ph]}}.

\bibitem{Aranda:2013gga}
A.~Aranda et~al., \emph{{Dirac neutrinos from flavor symmetry}},
  \MYhref[journalLinks]{http://dx.doi.org/10.1103/PhysRevD.89.033001}{Phys.
  Rev.
  }\MYhref[journalLinks]{http://dx.doi.org/10.1103/PhysRevD.89.033001}{\textbf{D89}
  (2014) 3 033001},
  \MYhref[eprintLinks]{http://arxiv.org/abs/1307.3553}{{\ttfamily
  arXiv:1307.3553 [hep-ph]}}.

\bibitem{Bonilla:2016diq}
C.~Bonilla, E.~Ma, E.~Peinado and J.~W.~F. Valle, \emph{{Two-loop Dirac
  neutrino mass and WIMP dark matter}},
  \MYhref[journalLinks]{http://dx.doi.org/10.1016/j.physletb.2016.09.027}{Phys.
  Lett.
  }\MYhref[journalLinks]{http://dx.doi.org/10.1016/j.physletb.2016.09.027}{\textbf{B762}
  (2016) 214--218},
  \MYhref[eprintLinks]{http://arxiv.org/abs/1607.03931}{{\ttfamily
  arXiv:1607.03931 [hep-ph]}}.

\bibitem{Ma:2016mwh}
E.~Ma and O.~Popov, \emph{{Pathways to Naturally Small Dirac Neutrino Masses}},
  \MYhref[journalLinks]{http://dx.doi.org/10.1016/j.physletb.2016.11.027}{Phys.
  Lett.
  }\MYhref[journalLinks]{http://dx.doi.org/10.1016/j.physletb.2016.11.027}{\textbf{B764}
  (2017) 142--144},
  \MYhref[eprintLinks]{http://arxiv.org/abs/1609.02538}{{\ttfamily
  arXiv:1609.02538 [hep-ph]}}.

\bibitem{Ma:2015raa}
E.~Ma and R.~Srivastava, \emph{{Dirac or inverse seesaw neutrino masses from
  gauged $B–L$ symmetry}},
  \MYhref[journalLinks]{http://dx.doi.org/10.1142/S0217732315300207}{Mod. Phys.
  Lett.
  }\MYhref[journalLinks]{http://dx.doi.org/10.1142/S0217732315300207}{\textbf{A30}
  (2015) 26 1530020},
  \MYhref[eprintLinks]{http://arxiv.org/abs/1504.00111}{{\ttfamily
  arXiv:1504.00111 [hep-ph]}}.

\bibitem{Ma:2015mjd}
E.~Ma, N.~Pollard, R.~Srivastava and M.~Zakeri, \emph{{Gauge $B-L$ Model with
  Residual $Z_3$ Symmetry}},
  \MYhref[journalLinks]{http://dx.doi.org/10.1016/j.physletb.2015.09.010}{Phys.
  Lett.
  }\MYhref[journalLinks]{http://dx.doi.org/10.1016/j.physletb.2015.09.010}{\textbf{B750}
  (2015) 135--138},
  \MYhref[eprintLinks]{http://arxiv.org/abs/1507.03943}{{\ttfamily
  arXiv:1507.03943 [hep-ph]}}.

\bibitem{Bonilla:2016zef}
C.~Bonilla and J.~W.~F. Valle, \emph{{Naturally light neutrinos in $Diracon$
  model}},
  \MYhref[journalLinks]{http://dx.doi.org/10.1016/j.physletb.2016.09.022}{Phys.
  Lett.
  }\MYhref[journalLinks]{http://dx.doi.org/10.1016/j.physletb.2016.09.022}{\textbf{B762}
  (2016) 162--165},
  \MYhref[eprintLinks]{http://arxiv.org/abs/1605.08362}{{\ttfamily
  arXiv:1605.08362 [hep-ph]}}.

\bibitem{Valle:2016kyz}
J.~W.~F. Valle and C.~A. Vaquera-Araujo, \emph{{Dynamical seesaw mechanism for
  Dirac neutrinos}},
  \MYhref[journalLinks]{http://dx.doi.org/10.1016/j.physletb.2016.02.031}{Phys.
  Lett.
  }\MYhref[journalLinks]{http://dx.doi.org/10.1016/j.physletb.2016.02.031}{\textbf{B755}
  (2016) 363--366},
  \MYhref[eprintLinks]{http://arxiv.org/abs/1601.05237}{{\ttfamily
  arXiv:1601.05237 [hep-ph]}}.

\bibitem{Abbas:2016qbl}
G.~Abbas, M.~Z. Abyaneh and R.~Srivastava, \emph{{Precise predictions for Dirac
  neutrino mixing}},
  \MYhref[journalLinks]{http://dx.doi.org/10.1103/PhysRevD.95.075005}{Phys.
  Rev.
  }\MYhref[journalLinks]{http://dx.doi.org/10.1103/PhysRevD.95.075005}{\textbf{D95}
  (2017) 7 075005},
  \MYhref[eprintLinks]{http://arxiv.org/abs/1609.03886}{{\ttfamily
  arXiv:1609.03886 [hep-ph]}}.

\bibitem{Abbas:2013uqh}
G.~Abbas, S.~Gupta, G.~Rajasekaran and R.~Srivastava, \emph{{High Scale Mixing
  Unification for Dirac Neutrinos}},
  \MYhref[journalLinks]{http://dx.doi.org/10.1103/PhysRevD.91.111301}{Phys.
  Rev.
  }\MYhref[journalLinks]{http://dx.doi.org/10.1103/PhysRevD.91.111301}{\textbf{D91}
  (2015) 11 111301},
  \MYhref[eprintLinks]{http://arxiv.org/abs/1312.7384}{{\ttfamily
  arXiv:1312.7384 [hep-ph]}}.

\bibitem{Wang:2016lve}
W.~Wang and Z.-L. Han,
  \MYhref[journalLinks]{http://dx.doi.org/10.1007/JHEP04(2017)166}{\emph{{Naturally
  Small Dirac Neutrino Mass with Intermediate $SU(2)_{L}$ Multiplet Fields}}
  }\MYhref[journalLinks]{http://dx.doi.org/10.1007/JHEP04(2017)166}{ (2016)},
  \MYhref[eprintLinks]{http://arxiv.org/abs/1611.03240}{{\ttfamily
  arXiv:1611.03240 [hep-ph]}}.

\bibitem{Wang:2017mcy}
W.~Wang, R.~Wang, Z.-L. Han and J.-Z. Han, \emph{{The $B-L$ Scotogenic Models
  for Dirac Neutrino Masses}}  (2017),
  \MYhref[eprintLinks]{http://arxiv.org/abs/1705.00414}{{\ttfamily
  arXiv:1705.00414 [hep-ph]}}.

\bibitem{Okada:2014vla}
H.~Okada, \emph{{Two loop Induced Dirac Neutrino Model and Dark Matters with
  Global $U(1)'$ Symmetry}}  (2014),
  \MYhref[eprintLinks]{http://arxiv.org/abs/1404.0280}{{\ttfamily
  arXiv:1404.0280 [hep-ph]}}.

\bibitem{Borah:2016zbd}
D.~Borah and A.~Dasgupta, \emph{{Common Origin of Neutrino Mass, Dark Matter
  and Dirac Leptogenesis}},
  \MYhref[journalLinks]{http://dx.doi.org/10.1088/1475-7516/2016/12/034}{JCAP
  }\MYhref[journalLinks]{http://dx.doi.org/10.1088/1475-7516/2016/12/034}{\textbf{1612}
  (2016) 12 034},
  \MYhref[eprintLinks]{http://arxiv.org/abs/1608.03872}{{\ttfamily
  arXiv:1608.03872 [hep-ph]}}.

\bibitem{Borah:2017leo}
D.~Borah and A.~Dasgupta, \emph{{Naturally Light Dirac Neutrino in Left-Right
  Symmetric Model}}  (2017),
  \MYhref[eprintLinks]{http://arxiv.org/abs/1702.02877}{{\ttfamily
  arXiv:1702.02877 [hep-ph]}}.

\bibitem{Schechter:1981cv}
J.~Schechter and J.~W.~F. Valle, \emph{{Neutrino Decay and Spontaneous
  Violation of Lepton Number}},
  \MYhref[journalLinks]{http://dx.doi.org/10.1103/PhysRevD.25.774}{Phys. Rev.
  }\MYhref[journalLinks]{http://dx.doi.org/10.1103/PhysRevD.25.774}{\textbf{D25}
  (1982) 774}.

\bibitem{Ma:2015pma}
E.~Ma, \emph{{Transformative $A_4$ mixing of neutrinos with CP violation}},
  \MYhref[journalLinks]{http://dx.doi.org/10.1103/PhysRevD.92.051301}{Phys.
  Rev.
  }\MYhref[journalLinks]{http://dx.doi.org/10.1103/PhysRevD.92.051301}{\textbf{D92}
  (2015) 5 051301},
  \MYhref[eprintLinks]{http://arxiv.org/abs/1504.02086}{{\ttfamily
  arXiv:1504.02086 [hep-ph]}}.

\bibitem{Forero:2014bxa}
D.~Forero, M.~Tortola and J.~W.~F. Valle, \emph{{Neutrino oscillations
  refitted}},
  \MYhref[journalLinks]{http://dx.doi.org/10.1103/PhysRevD.90.093006}{Phys.Rev.
  }\MYhref[journalLinks]{http://dx.doi.org/10.1103/PhysRevD.90.093006}{\textbf{D90}
  (2014) 9 093006},
  \MYhref[eprintLinks]{http://arxiv.org/abs/1405.7540}{{\ttfamily
  arXiv:1405.7540 [hep-ph]}}.

\bibitem{Schechter:1980gr}
J.~Schechter and J.~W.~F. Valle, \emph{{Neutrino Masses in SU(2) x U(1)
  Theories}},
  \MYhref[journalLinks]{http://dx.doi.org/10.1103/PhysRevD.22.2227}{Phys.Rev.
  }\MYhref[journalLinks]{http://dx.doi.org/10.1103/PhysRevD.22.2227}{\textbf{D22}
  (1980) 2227}.

\bibitem{Morisi:2009sc}
S.~Morisi and E.~Peinado, \emph{{An A4 model for lepton masses and mixings}},
  Phys. Rev. \textbf{D80} (2009) 113011,
  \MYhref[eprintLinks]{http://arxiv.org/abs/0910.4389}{{\ttfamily
  arXiv:0910.4389 [hep-ph]}}.

\bibitem{Xing:2011aa}
Z.-z. Xing, H.~Zhang and S.~Zhou, \emph{{Impacts of the Higgs mass on vacuum
  stability, running fermion masses and two-body Higgs decays}},
  \MYhref[journalLinks]{http://dx.doi.org/10.1103/PhysRevD.86.013013}{Phys.
  Rev.
  }\MYhref[journalLinks]{http://dx.doi.org/10.1103/PhysRevD.86.013013}{\textbf{D86}
  (2012) 013013},
  \MYhref[eprintLinks]{http://arxiv.org/abs/1112.3112}{{\ttfamily
  arXiv:1112.3112 [hep-ph]}}.

\bibitem{Bonilla:2016sgx}
C.~Bonilla, M.~Nebot, R.~Srivastava and J.~W.~F. Valle, \emph{{Flavor physics
  scenario for the 750 GeV diphoton anomaly}},
  \MYhref[journalLinks]{http://dx.doi.org/10.1103/PhysRevD.93.073009}{Phys.
  Rev.
  }\MYhref[journalLinks]{http://dx.doi.org/10.1103/PhysRevD.93.073009}{\textbf{D93}
  (2016) 7 073009},
  \MYhref[eprintLinks]{http://arxiv.org/abs/1602.08092}{{\ttfamily
  arXiv:1602.08092 [hep-ph]}}.

\bibitem{Arnold:2017bnh}
R.~Arnold et~al., \emph{{Search for neutrinoless quadruple-$\beta$ decay of
  $^{150}$Nd with the NEMO-3 detector}}  (2017),
  \MYhref[eprintLinks]{http://arxiv.org/abs/1705.08847}{{\ttfamily
  arXiv:1705.08847 [hep-ex]}}.

\bibitem{Binosi:2003yf}
D.~Binosi and L.~Theussl, \emph{{JaxoDraw: A Graphical user interface for
  drawing Feynman diagrams}},
  \MYhref[journalLinks]{http://dx.doi.org/10.1016/j.cpc.2004.05.001}{Comput.
  Phys. Commun.
  }\MYhref[journalLinks]{http://dx.doi.org/10.1016/j.cpc.2004.05.001}{\textbf{161}
  (2004) 76--86},
  \MYhref[eprintLinks]{http://arxiv.org/abs/hep-ph/0309015}{{\ttfamily
  arXiv:hep-ph/0309015 [hep-ph]}}.

\end{thebibliography}

\end{document}